\documentstyle[aps,prb,epsf,floats]{revtex}

\newcommand{\ybco}{YBa$_2$Cu$_3$O$_{6.35}$}
\newcommand{\lsco}{La$_{2-x}$Sr$_x$CuO$_4$}
\newcommand{\bscco}{Bi$_2$Sr$_2$CaCu$_2$O$_{8+\delta}$}

\begin{document}
\twocolumn[\hsize\textwidth\columnwidth\hsize\csname@twocolumnfalse%
\endcsname

\title{
Superconducting charge-ordered states in cuprates
}

\author{Matthias Vojta}
\address{
Theoretische Physik III, Elektronische Korrelationen und
Magnetismus, Universit\"at Augsburg, 86135 Augsburg, Germany
}
\date{July 10, 2002}
\maketitle

\begin{abstract}
Motivated by recent neutron scattering and scanning tunneling microscopy (STM)
experiments on cuprate superconductors, we discuss charge-ordered states,
in particular with two-dimensional charge modulation patterns, co-existing
with superconductivity.
We extend previous studies of a large-$N$ mean-field formulation of the $t-J$ model.
In addition to bond-centered superconducting stripe states at low doping,
we find checkerboard-modulated superconducting states which are favorable
in an intermediate doping interval.
We also analyze the energy dependence of the Fourier component
of the local density of states at the ordering wavevector for several
possible modulation patterns, and compare with STM results.
\end{abstract}
]

\section{Introduction}

A series of recent experiments have highlighted the importance of
spin and charge ordering tendencies in the cuprate superconductors.
Static stripe order has been established in Nd-doped \lsco \cite{lsco} which appears
to co-exist with superconductivity at very low temperatures.
Recently, superconducting samples of underdoped \ybco \cite{mook}
and nearly optimally doped \bscco \cite{kapitul} with $T_c$ up to 90 K
have been found to show signatures of charge ordering.
In addition, a variety of compounds display strong dynamic spin
and charge fluctuations, which are enhanced by the application of
moderate magnetic fields \cite{seamus,lake2,boris}.
This emphasizes that even materials, where no static order can
be observed in the absence of an external field, are in close
proximity to a critical point where spin and/or charge
order is established in the superconducting state.
Theories based on this assumption \cite{sy,so5,demler}
successfully explain a number of
NMR \cite{imai} and neutron scattering \cite{lake2,boris,aeppli}
experiments.

In this paper, we will focus on superconducting states with
static charge order, but dynamic spin fluctuations -- this appears
to be realized in the experiments of Ref.~\onlinecite{mook}.
Such states are found upon doping paramagnetic Mott insulators
on the square lattice \cite{vs,vzs}.
The undoped quantum paramagnet has broken translational symmetry associated
with spontaneous bond charge (or spin-Peierls) order;
at small carrier concentration, $\delta$, this order persists and
co-exists with anisotropic superconductivity;
a $d$-wave superconductor with full square lattice symmetry appears
above a critical $\delta$.
[``Charge order'' is defined very generally as spatial modulation in any
SU(2)-invariant observables, such as local density of states (LDOS) per {\em site},
or kinetic exchange energy per lattice {\em bond};
the modulation in the total site charge density can be small
due to long-range Coulomb interactions.]
A particular feature of the superconducting charge-ordered states found in Refs.~\onlinecite{vs,vzs}
is that the modulation is bond-centered and its real-space period $p$ always
takes even integer values (in units of the Cu lattice spacing);
this is in contrast to the continuous doping evolution of the ordering
wavevector, $1/p$, usually assumed in the so-called ``Yamada plot'' \cite{yamada}.
Interestingly, the experimental results of Refs.~\onlinecite{mook,kapitul,seamus}
appear to be remarkably well described by the states proposed in
Refs.~\onlinecite{vs,vzs}, as has also been discussed in recent work \cite{damdem,chenhu}
which appeared while this paper was being completed:
Neutron scattering on underdoped \ybco \cite{mook} shows charge order with a real-space
period of 8 lattice sites, whereas \bscco\ close to optimal doping
displays $p\!=\!4$ modulation \cite{kapitul,seamus};
furthermore the STM data indicate an even-period modulation not only
in the site charge density, but also in the bond kinetic energy and perhaps
the bond pairing amplitude~\cite{damdem}.

We note that numerical studies of the $t-J$ model \cite{white} have observed
bulk charge order with period $p\!=\!4$ at doping level 1/8;
and paired hole states with different types of charge order 
in both insulators and superconductors have been discussed
elsewhere \cite{jan,ps,assa}.

In the past, most theoretical work has been focussed on states with
one-dimensional (1d) charge modulation, often referred to as stripes.
However, recent STM experiments \cite{kapitul,seamus} indicate LDOS
modulations in {\em both} $x$ and $y$ directions in a single CuO$_2$
plane (although a small anisotropy is observed).
Theoretically, charge density wave (CDW) fluctuations are expected in
both directions \cite{vzs} on the {\em disordered} side of the charge ordering transition
(if the system has no intrinsic lattice anisotropy).
Moving to the ordered side, it depends on microscopics whether CDW order in one
or in two directions condenses, leading to stripe-like or two-dimensional (2d)
modulations, respectively.

The purpose of this paper is twofold:
In Sec.~\ref{sec:mf}, we re-examine the mean-field theory of Refs.~\onlinecite{vs,vzs},
to investigate the possible existence of and the mechanism leading to superconducting
charge-ordered states with 2d charge modulation.
In Sec.~\ref{sec:ldos}, we turn to a detailed discussion of the charge modulation
pattern, by calculating the energy dependence of the LDOS Fourier component at
the ordering wavevector and comparing it with the measurements of
Ref.~\onlinecite{kapitul}.

\section{Mean-field theory}
\label{sec:mf}

We start with re-analyzing the large-$N$ theory of Refs.~\onlinecite{vs,vzs},
which provides a microscopic description of doping mobile charge carriers
into a paramagnetic Mott insulator \cite{pwa}.
We consider an extended $t-J$ Hamiltonian for
fermions, $c_{i \alpha}$, on the sites, $i$, of a square
lattice with spin $\alpha=1 \ldots 2N$
($N=1$ is the physical value):
\begin{eqnarray}
{\cal H}_{tJV} = \sum_{i > j} && \left[ -\frac{t_{ij}}{N} c_{i \alpha}^{\dagger} c_{j
\alpha} + {\rm H.c.}  + \frac{V_{ij}}{N} n_i n_j \right. \nonumber \\
&&~~~~~~~~+\left. \frac{J_{ij}}{N} \left( {\bf S}_i \cdot {\bf S}_j -
\frac{n_i n_j}{4N} \right) \right].
\label{e1}
\end{eqnarray}
Here $n_i = c_{i \alpha}^{\dagger} c_{i \alpha}$ is the on-site
charge density, and the spin operators ${\bf S}_i$ are fermion
bilinears times the traceless generators of ${\rm Sp} (2N)$.
For most of the following, the fermion hopping, $t_{ij}$, and
exchange, $J_{ij}$, will be restricted to nearest-neighbor terms,
$t$ and $J$; for the detailed comparison with the Ref.~\onlinecite{kapitul}
we will introduce 2nd neighbor hopping, $t'$.
The electronic Coulomb interaction is represented by the
on-site constraint $n_i \leq N$ and the off-site repulsive interactions
$V_{ij} = V/|r_{ij}|$.
The average doping $\delta$ is fixed by
$\sum_i \langle n_i \rangle = N N_s (1-\delta)$,
where $N_s$ is the number of lattice sites.
To proceed, we represent the spins by auxiliary fermions $f_{i\alpha}$
and the holes by spinless bosons $b_i$, such that the physical
electrons $c_{i\alpha} = b_i^\dagger f_{i\alpha}$, and the
necessary Hilbert space constraint is implemented by Lagrange
multipliers $\lambda_i$.
Via a Hubbard-Stratonovich decoupling of the antiferromagnetic
interaction we introduce link fields $Q_{ij}$, defined on the
bonds of the square lattice.
After taking the limit $N\to\infty$, the slave bosons $b_i$ condense,
$\langle b_i\rangle = \sqrt{N}b_i$,
the $Q_{ij}$ and $\lambda_i$ take static saddle-point values, and we are
left with a bilinear Hamiltonian which can be diagonalized by
a Bogoliubov transformation.
At the saddle point, the slave boson amplitudes fulfill
$\sum_i b_i^2 = N_s \delta$, and the link fields are given by
$N Q_{ij}=\langle {\cal J}^{\alpha\beta} f_{i\alpha}^\dagger f_{j\beta}^\dagger \rangle$,
where ${\cal J}^{\alpha\beta}$ is the antisymmetric ${\rm Sp}(2N)$ tensor;
for further details see Ref.~\onlinecite{vzs}.

Various ground states obtained from the numerical solution of the above mean-field
equations have been discussed in Refs.~\onlinecite{vs,vzs}.
At $\delta=0$ the ground state is a fully dimerized, insulating spin-Peierls
state, i.e., it has period-2 bond charge order.
At low doping $\delta$, the bare large-$N$ $t-J$ model tends to phase separation,
and the inclusion of moderate Coulomb interaction, $V$, leads to the formation
of bond-centered, superconducting stripes with a 1d
charge modulation.
Large doping destroys charge order and leads to a pure $d$-wave superconducting ground state,
in this regime the large-$N$ approach reduces to the usual BCS mean-field
theory with renormalized hopping matrix elements.

Motivated by the STM results of Refs.~\onlinecite{kapitul,seamus}
we have searched for additional saddle-point solutions with 2d
charge modulation; we have restricted the attention to states in which
the charge distribution respects the 90 degree rotation symmetry of the
lattice.
Interestingly, there are several such saddle points which were overlooked
in Ref.~\onlinecite{vzs}.
In most of the low and intermediate doping region the states with 1d and 2d
modulation are close in energy;
at small doping the 1d stripe-like states are preferred, whereas the
2d checkerboard-likes states are lower in energy in a certain interval
of intermediate doping and Coulomb repulsion.

\begin{figure}
\epsfxsize=3.2in
\centerline{\epsffile{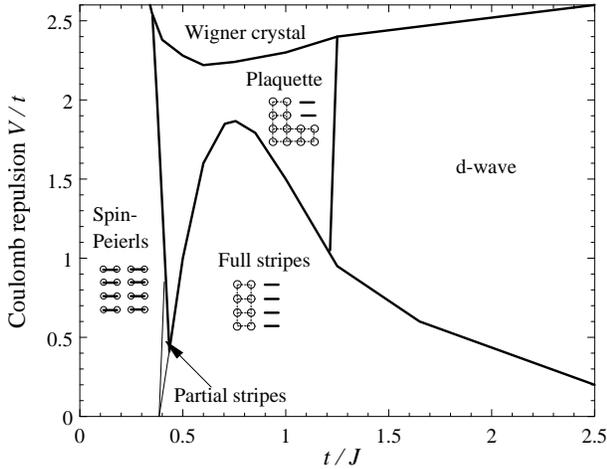}}
\caption{
Ground state phase diagram of the extended $t-J$ model
in the large-$N$ limit at doping $\delta\!=\!20\%$.
Thick (thin) lines indicate first- (second-) order
transitions.
All states except for the Wigner crystal have superconducting order.
The stripe and spin-Peierls phases show 1d charge modulation;
the plaquette phase has charge modulations with
real-space period 4 in both directions.
The circles indicate the spatial distribution of hole density,
the lines symbolize the strength of the bond variables
$Q_{ij}$.
``Full stripes'' refers to states where the charge modulation in the
large-$N$ limit is maximal, i.e., the hole density is zero in the
hole-poor regions, whereas the ``partial stripe'' states have a finite
hole density there.
}
\label{fig:pd}
\end{figure}

We have therefore concentrated on states with a real-space periodicity
$p\!=\!4$, leading to a unit cell of $4\times4$
sites \cite{kapitul}.
(Recall that bond-centered stripes with the spatial period pinned to four
sites were found over a rather large range of doping values in
Refs.~\onlinecite{vs,vzs}.)
The most favorable mean-field states with 2d charge modulation are
characterized by holes arranged in intersecting bond-centered stripes,
i.e., the hole concentration is large on 12 sites and zero on 4 sites,
see Fig.~\ref{fig:pd}.
The actual filling in the hole-rich regions is smaller than 1 and depends
on microscopic parameters; due to the strong pairing correlations
the 2d CDW states are good superconductors (with a $d$-wave-like pairing symmetry).
Such $4\times4$ plaquette states occur as large-$N$ ground states of
the Hamiltonian (\ref{e1}) for doping levels between approximately 13\% and 25\%.
A sample phase diagram for doping $\delta=20\%$ is shown in Fig.~\ref{fig:pd}
(see also Fig.~3 of Ref.~\onlinecite{vs}).
For small $t/J$ phase separation tendencies do not occur, therefore
the ground state is a doped spin-Peierls state with homogeneous site
charge distribution.
At intermediate $t/J$, frustrated phase separation leads either to stripes
or to plaquette states;
at larger $t/J$ the kinetic energy becomes dominant and weakens the phase
separation tendencies, consequently a homogeneous $d$-wave state is reached.
At very large Coulomb repulsion, the holes arrange into an insulating
Wigner crystal.

The occurence of plaquette-modulated CDW states in favor of stripes
can be understood as an interplay of exchange, kinetic, and Coulomb
energy as follows:
The exchange term prefers dimerization between spins on neighboring sites
and tends to expel holes -- this produces the overall dimer structure
of the $Q_{ij}$, 
and leads to a fraction of sites being undoped.
The kinetic energy is lowered by possible hopping processes, i.e.,
by neighboring sites with non-zero hole density.
Clearly, the exchange term prefers stripes, whereas the kinetic term
prefers plaquettes, where hopping in two directions is possible.
Now, the Coulomb energy of the 2d modulated state is significantly
lower than that of a stripe state,
because the charge inhomogeneity is smaller in the plaquette state
(which is also closer to a crystalline arrangement of charges).
From this discussion it is clear that at low doping, where the physics is dominated
by the exchange term, stripe states are preferred.
With increasing doping the kinetic energy becomes more and more important,
which leads to 2d modulated CDW states at moderate values of $V$,
before a homogeneous $d$-wave superconductor becomes the ground
state.

In all CDW states superconductivity competes with charge order.
In the stripe states superconductivity is very weak due to the strong
anisotropy: bulk superconductivity is established only by Cooper pair
tunneling between the stripes.
In contrast, the plaquette CDW states are much better superconductors,
due to the full 2d character of the charge distribution.
This trend is consistent with the low $T_c$ in Nd-doped \lsco \cite{lsco}
compared to apparently charge-ordered \ybco \cite{mook} and \bscco samples
\cite{kapitul}.

\section{Modulation in the local density of states}
\label{sec:ldos}

After having established the possible occurence of plaquette CDW states
in the large-$N$ theory for the $t-J$ model,
we turn to a detailed analysis of the corresponding STM signal.
Hoffman {\em et al.} \cite{seamus} have introduced a STM technique of
atomically resolved spectroscopic mapping, which allowed to detect
LDOS modulations around vortex cores with real-space period 4, i.e.,
at wavevectors ${\bf K}_x=(\pi/2,0)$ and ${\bf K}_y = (0,\pi/2)$.
Howald {\em et al.} \cite{kapitul} used this technique to map the energy
dependence, $\rho_{\bf K}(\omega)$, of the spatial Fourier component of
the LDOS at the ordering wavevectors ${\bf K}_{x,y}$.
This energy dependence has been recently discussed within a model
for the pinning of SDW/CDW fluctuations by inhomegeneities~\cite{pin},
and in an analysis of different patterns of translational symmetry breaking
in $d$-wave superconductors \cite{damdem}.

\begin{figure}
\epsfxsize=3in
\centerline{\epsffile{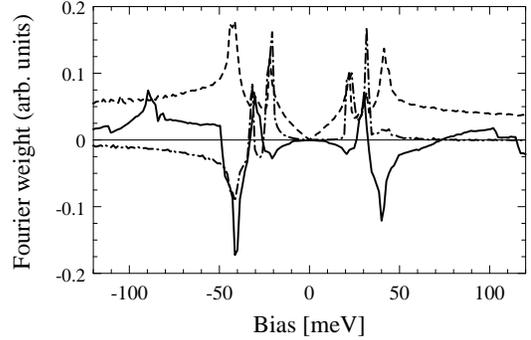}}
\caption{
Energy dependence $\rho_{\bf K}(\omega)$ of the Fourier component of the
LDOS at ${\bf K}=(\pi/2,0)$, obtained from diagonalizing
${\cal H}_{\rm BCS}+{\cal H}_{\rm mod}$ on a $4\times4$ unit cell.
Bulk parameter values are $t=0.15$ eV, $t'=-t/4$, 
doping $\delta=17\%$, and a gap size $\Delta_0=40$~meV.
The curves correspond to modulations as follows: solid -- site charge density,
dashed -- bond charge density, dash-dot: pairing amplitude.
The amplitude of the LDOS modulation is proportional to $V_0$ in ${\cal H}_{\rm mod}$;
here $V_0^{\rm site}=V_0^{\rm kin}=10$ meV, $V_0^{\rm pair}=4$ meV.
}
\label{fig:ldos1}
\end{figure}

For a comparison with the experimental situation~\cite{kapitul} we restrict
ourselves again to states with period-4 modulation; we will
employ hopping parameters $t$, $t'$ that yield a realistic
band structure.
Furthermore we have to keep in mind the shortcomings of the
large-$N$ theory: the precise location of the phase boundaries is not reliable,
and the theory underestimates fluctuations.
Therefore we will work with superconducting gap values close to the
experimentally observed ones, and discuss both self-consistent mean-field
solutions as well as states where translational symmetry breaking is imposed
by hand in the Hamitonian.

To obtain initial information about the possible forms of $\rho_{\bf K}(\omega)$
we start by considering $d$-wave superconductors with additional modulation
in one of the following quantities: site charge density, bond charge density (kinetic energy),
pairing amplitude.
Such an analysis has also been independently performed by Podolsky {\em et al.} \cite{damdem},
but here we are interested in 2d modulations and furthermore diagonalize
the mean-field Hamiltonian exactly for the $4\times4$ unit cell.
The Hamiltonian thus has the form ${\cal H}_{\rm BCS}+{\cal H}_{\rm mod}$,
where
\begin{equation}
{\cal H}_{\rm BCS} =
\sum_{\bf k} \epsilon_{\bf k} c_{{\bf k},\sigma}^\dagger c_{{\bf k},\sigma} +
\sum_{\bf k} \Delta_{\bf k} (c_{{\bf k},\uparrow}^\dagger c_{-{\bf k},\downarrow} + h.c.)
\label{hbcs}
\end{equation}
in standard notation, with $\Delta_{\bf k}=\Delta_0 (\cos k_x - \cos k_y)/2$;
${\cal H}_{\rm BCS}$ is equivalent to the ${\rm Sp}(2N)$ mean-field theory presented
above in the region where the large-$N$ ground state is a pure $d$-wave superconductor
(with the correspondence $\Delta_{ij} = J_{ij} Q_{ij}$ where $\Delta_{ij}$ is the
real-space Fourier transform of the energy gap $\Delta_{\bf k}$).

The modulation is introduced via ${\cal H}_{\rm mod}$:
for site CDW we add
${\cal H}_{\rm mod}^{\rm site} = V_0^{\rm site} \sum_i f({\bf R}_i) c_{i\sigma}^\dagger c_{i\sigma}$,
for bond CDW we have
${\cal H}_{\rm mod}^{\rm kin} = V_0^{\rm kin} \sum_{\langle ij\rangle} f[({\bf R}_i+{\bf R}_j)/2] c_{i\sigma}^\dagger c_{j\sigma}$,
and a pairing modulation is given by
${\cal H}_{\rm mod}^{\rm pair} = V_0^{\rm pair} \sum_{\langle ij\rangle} f[({\bf R}_i+{\bf R}_j)/2]
(c_{i\uparrow}^\dagger c_{j\downarrow}^\dagger + h.c.)$.
The function $f({\bf R})$ describes modulation strength and pattern, and we will concentrate on
2d bond-centered period-4 modulations with
$f({\bf R}) = [\cos(\pi R_x/2 + \pi/4) + \cos(\pi R_y/2 + \pi/4)] / 2$.

Diagonalization of ${\cal H}_{\rm BCS}+{\cal H}_{\rm mod}$ yields the local density
of states for each site of the unit cell, from which we find $\rho_{\bf K}(\omega)$
by Fourier transformation
(the real-space origin is chosen such that $\rho_{\bf K}(\omega)$ is real);
note that ${\bf K}=(\pi/2,0)$ and $(0,\pi/2)$ are
equivalent with the above choice of $f({\bf R})$.
Results for $\rho_{\bf K}(\omega)$ are shown in Fig.~\ref{fig:ldos1} for the
three modulation cases listed above.
If we compare the curves in Fig.~\ref{fig:ldos1} with the STM result in
Fig.~3 of Ref.~\onlinecite{kapitul}, which shows a peak in the magnitude
of $\rho_{\bf K}(\omega)$ at subgap energies $|\omega|/\Delta_0\approx 2/3$,
it is clear that the experiments are not well described by a site charge
modulation alone.
In contrast, our result for a modulation in the pairing amplitude comes
closest to the curves of Ref.~\onlinecite{kapitul}.
We note that our results in Fig.~\ref{fig:ldos1} are somewhat different
from the ones of Ref.~\onlinecite{damdem}, this may be
due to the 2d character of the modulation considered
here and due to the approximations employed in Ref.~\onlinecite{damdem}.

\begin{figure}
\epsfxsize=3in
\centerline{\epsffile{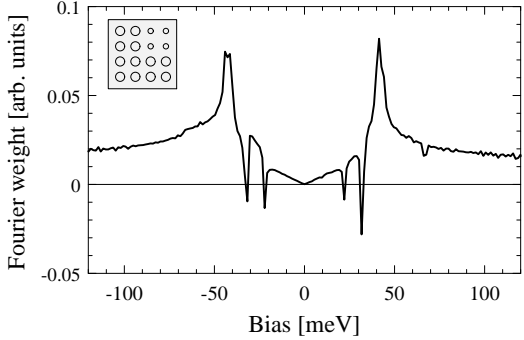}}
\caption{
As in Fig.~\protect\ref{fig:ldos1}, but for a state with plaquette
site charge modulation as indicated in the inset ($V_0^{\rm site}=5$~meV),
and self-consistently determined pair fields $\Delta_{ij}$, using $J=70$~meV.
Note that LDOS modulations of similar magnitude occur at wavevectors
$\pm{\bf K}_x\pm{\bf K}_y$.
}
\label{fig:ldos2}
\end{figure}

It is clear that the experimentally realized CDW state will have modulations
in all quantities invariant under spin rotations and time reversal.
This can -- at least in part -- be captured by a self-consistent solution of
the mean-field equations.
A natural candidate is given by the plaquette state found as large-$N$ ground state
above (Fig.~\ref{fig:pd}).
Thus, we employ the Hamiltonian ${\cal H}_{\rm BCS}+{\cal H}_{\rm mod}^{\rm site}$,
where the pair-field $\Delta$ in ${\cal H}_{\rm BCS}$ (\ref{hbcs}) is determined
self-consistently from
$\Delta_{ij}= J_{ij} \langle {\cal J}^{\alpha\beta} c_{i\alpha}^\dagger c_{j\beta}^\dagger\rangle$,
and ${\cal H}_{\rm mod}^{\rm site}$ imposes a weak modulation of the site charge density
as shown in Fig.~\ref{fig:ldos2}
(the strong modulation found in the large-$N$ limit will certainly be weakened
by fluctuation corrections beyond the large-$N$ theory).
Fig.~\ref{fig:ldos2} displays a corresponding LDOS modulation $\rho_{\bf K}(\omega)$.
The agreement with the available experimental data \cite{kapitul} is
not satisfying, in particular $\rho_{\bf K}(\omega)$ does not show a large peak at
energies below the bulk superconducting gap.

\begin{figure}
\epsfxsize=3in
\centerline{\epsffile{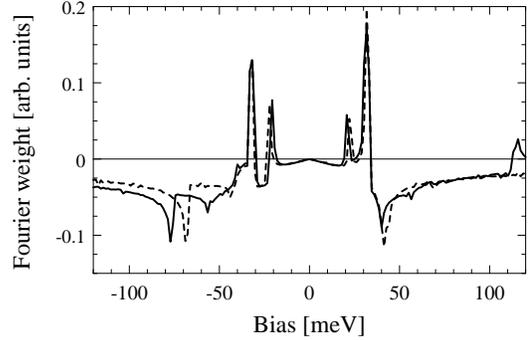}}
\caption{
As in Fig.~\protect\ref{fig:ldos1}, but for a model with plaquette
modulation in both site charge density and exchange interaction, and
self-consistently determined pair fields $\Delta_{ij}$,
using $V_0^{\rm site}=10$ meV, $J_0=70$ meV, $V_0^J=3.5$ meV.
The dashed curve corresponds to a band structure with
$t'=-t/3$ to demonstrate the robustness of the result.
}
\label{fig:ldos3}
\end{figure}

This fact and the results in Fig.~\ref{fig:ldos1} led us to consider an additional
effect not captured in the mean-field calculations:
On general symmetry grounds, a static charge modulation will lead to a real-space
modulation in the effective {\em pairing interaction}, because the CDW influences
the local spin fluctuation spectrum.
On the mean-field level, this can be phenomenologically accounted for by a modulation
in the exchange interaction $J$.
Therefore, we have studied self-consistent solutions of the mean-field
theory, ${\cal H}_{\rm BCS}+{\cal H}_{\rm mod}^{\rm site}$, as above, but
in addition to a weak site charge modulation in ${\cal H}_{\rm mod}^{\rm site}$
we imposed a modulation of the $J_{ij}$ exchange interaction,
$J_{ij} = J_0 + V_0^J f[({\bf R}_i+{\bf R}_j)/2]$ for nearest neighbor sites $i$ and $j$,
which leads to corresponding modulations in both the pair fields
$\Delta_{ij}$ and the bond charge density (kinetic energy).
Results for the LDOS Fourier component $\rho_{\bf K}(\omega)$ are shown
in Fig.~\ref{fig:ldos3}, with a rather good agreement with
the experiments of Ref.~\onlinecite{kapitul}.

\section{Conclusions}

Summarizing, we have studied superconducting charge-ordered states
of doped Mott insulators.
Within a large-$N$ theory we have established that ground states with
2d charge modulation can occur
at intermediate doping where they are preferred over stripes.
By analyzing the energy dependence of the LDOS modulation as observed
in STM, we have found the data of Ref.~\onlinecite{kapitul} to be well
described by combined modulations in charge density as well as exchange
and pairing energy, caused by a modulation of the pairing interaction.


\acknowledgements

The author thanks S. Davis, E. Demler, A. Polkovnikov, and J. Zaanen
for valuable discussions,
and in particular S. Sachdev 
for suggesting a search for plaquette states within
the formalism of Refs.~\onlinecite{vs,vzs}.
This research was supported by the DFG through SFB 484.



\end{document}